\documentclass[pra,preprint,showpacs,preprintnumbers,amsmath,amssymb]{revtex4}
\usepackage{epsf,graphicx}% Include figure files

\newcommand{\ket}[1]{\mathop{\left| #1 \right\rangle}\nolimits}
\newcommand{\bra}[1]{\mathop{\left\langle #1 \right|}\nolimits}
\newcommand{\brawavy}[1]{\mathop{\left(#1\right|}\nolimits}
\newcommand{\ketwavy}[1]{\mathop{\left|#1\right)}\nolimits}
\newcommand{\braketwavy}[2]{\mathop{(#1\left.\right|#2
)}\nolimits}

\begin{document}
\draft

\title{Generalized quantum measurements. Part II: Partially-destructive
quantum measurements in finite-dimensional Hilbert spaces}

\author{Boris A.\ Grishanin}\email{grishan@comsim1.phys.msu.ru}
\author{Victor N. Zadkov}
\affiliation{International Laser Center and Faculty of Physics\\
M.\ V.\ Lomonosov Moscow State University, 119899 Moscow, Russia}

\date{June 6, 2005}

\begin{abstract}
A concept of the generalized quantum measurement is introduced as the
transformation, which establishes a correspondence between the initial states
of the object system and final states of the object--measuring device (meter)
system with the help of a classical informational index, unambiguously linked
to the classically compatible set of states of the object--meter system. It is
shown that the generalized measurement covers all the key known quantum
measurement concepts---standard projective, entangling, fuzzy and the
generalized measurement with the partial or complete destruction of the initial
information contained in the object. A special class of partially-destructive
measurements that map the continual set of the states in finite-dimensional
quantum systems to that one of the infinite-dimensional quantum systems is
considered. Their informational essence and some information characteristics
are discussed in detail.
\end{abstract}

\maketitle

\section{Introduction\label{introduction}}

Recent progress in developing sophisticated modern methods for engineering of
quantum information \cite{bowmeester} caused significant changes in the
standard (quasiclassical) concept of the quantum measurement. From the early
days of quantum mechanics, the procedure of quantum measurement is interpreted
as setting a correspondence between the eigenstates of a measurable variable of
the object quantum system (in the following, simply the object) and the
indicator variable of the \emph{quasiclassical} measuring device (in the
following, simply the meter), which measures the state of the object system
\cite{Neumann,sudbery}. Nowadays, this concept of the quantum measurement has
been changed towards more general concept that is setting the respected
correspondence between the object and \emph{quantum} meter in an essentially
\emph{quantum form}, which generally includes an entanglement in the
object--meter system \cite{PRA2003} (the so called \emph{entangling}
measurement).

In the frame of this concept, it is natural to introduce the class of the
``soft" measurements as the measurements accuracy of which is limited with the
internal quantum uncertainty of the states of the indicator variable used in
the measurement \cite{JETP2004}. It presents the most basic class of a more
general definition of a fuzzy measurement \cite{mensky}. In the case of a
continual-valued measurable variable, studying such class of measurements
reveals the information essence of the internal quantum uncertainty of the
nonorthogonal quantum states of the indicator, so that it is described not in
the terms of the fluctuations of the physical variables (compare with the
Heisenberg inequality) widely operated by physicists, but directly in the terms
of the quantum states, i.e., irrespective of the values of the physical
variables themselves.

For the classical standard measurement, one can easily contribute the
requirement of the absolute accuracy of the measurement result with the
requirement of the absolute absence of the perturbations in the measurable
system during the measuring procedure. By contrast, in the generalized concept
of the quantum measurement both these requirements cannot be fulfilled simply
due to the specific properties of the set of quantum states forming a linear
space. Any interaction with a quantum system inevitably changes at least a part
of its possible quantum states. Being ever significant in terms of the Hilbert
space, these changes  are not take into account when we map onto the classical
system the quantum algebra of the observable variables with the use of the
classical limit $\hbar\to0$.

For a quantum system, the requirement for the absolute absence of the
perturbations can be fulfilled only with respect to the measurable variable
selecting therefore a special class of the so called \emph{nondemolition}
measurements. Due to the uniqueness taken into account by the no-cloning
principle \cite{RTE02}) of the entire quantum information the nondemolition
measurement completely destroys the coherent, i.e., essentially quantum,
information in the initial state of the object \cite{PRA2003}, which in the
case of completely coherent measurement is distributed among the object and the
meter and does not exist in the separate systems of the object and the meter.

Selecting the subclass of the soft nondemolition quantum measurements allows
partial preserving of the coherent information in the object. On contrary,
transmission of the major portion of information towards the meter can be
realized only within the class of the so called \emph{desctructive}
(demolition) measurement, when the state of the object is inevitably perturbed.
A limiting case of maximally destructive measurement, which entirely destroys
the initial information can be illustrated with the totally coherent transition
of the excitation from one oscillator to another one that belong to a set of
coupled oscillators. Another illustrating example can be given by considering a
purely non-coherent measurement of the state of a two-level atom with the help
of detection of the irradiated photon.

In this paper, we consider a more general class of measurements, which by
contrast with the class of the soft measurements defined in
Ref.~\onlinecite{JETP2004} allows perturbation of the initial state along any
variable. Respectively, we analyze the most interesting, in respect to the
qualitative content of the mapped information, case of nonselected mapping of
all states of the Hilbert space $H$ of a quantum system onto the classically
distinguishable eigen states of a continuous variable of another more complex
quantum system with the Hilbert space $L_2(H)$.

This class of quantum measurements, along with showing up the potential
resources of the measuring transformation of this kind for developing new
methods of engineering of quantum information, is also interested from the
fundamental point of view of qualitative interpretation of quantum theory. It
helps to exposure the most general relationships between the physical content
of the transformations applied to a quantum system and the classical
information contained in the values of the information index, which sets the
unique correspondence between the initial quantum states and the quantum states
after the measurement.

\section{Definition of the indexing destructive measurement and its correspondence
with the generalized measurement\label{Definitions}}

We will start with the isometric transformation of the form
\begin{equation}\label{V}
{\cal V}=\sum\limits_\alpha\sqrt{\nu_\alpha}\ket\alpha_{AB}\brawavy\alpha_A
\end{equation}

\noindent from Hilbert space $H_A$ of the object $A$ onto the space $H_A\otimes H_B$
of the bipartite system object--meter $A+B$. Here vectors $\ket\alpha_{AB}$ define
the {\em orthogonal} basis in the Hilbert space of the bipartite system
object--meter, indexed by the values $\alpha$ of an indicator variable. This basis
allows new representation of the initial quantum information that can be measured in
general case with the help of \emph{nonorthogonal} ``probe'' states
$\brawavy\alpha_A$; the set of positive numbers $\nu_\alpha$ characterizes the
repetition factor of the elementary maps $\ketwavy\alpha_A\to\ket\alpha_{AB}$, of
which the resulted transformation ${\cal V}$ is constructed as the coherent (i.e.,
depending on the phases of the wave functions) superposition of the respective
generalized projectors.

The relation
\begin{equation}\label{norm}
 {\cal V}^+{\cal V}\equiv\sum_\alpha\nu_\alpha\ketwavy\alpha_A\brawavy\alpha_A=\hat
 I_A.
\end{equation}

\noindent assures the isometric property of the transformation. It admittedly
can be fulfilled if the set of mapped states $\ketwavy\alpha_A$ is a set of
orthogonal bases, randomly rotated with respect to each other. Specifically,
for $N$ equally represented bases in a $D$-dimensional space we have
$\nu_\alpha=1/ND$.

The transformation (\ref{V}) is a generalized modification of the canonical
representation of the isometric mapping ${\cal V}=\sum\ket k_C \bra k_A$ as the
transformation of the entire orthogonal set in $H_A$ into an orthogonal set in
an arbitrary space $H_C$. This transformation, first of all, concretizes the
structure of the mapping space as the space of the states of the bipartite
system object--meter, $A+B$. Secondly, it uses in general case an overfull set
of states $\ketwavy\alpha$ for the representation of the set of the initial
states.

The physical meaning of consideration of the isometric mapping of $A$ onto $A+B$ is
that it can always be redefined up to the unitary transformation $U$ in the
bipartite system $A+B$, which corresponds to the transformation ${\cal V}$ at the
fixed initial state $\ket0_B$ of the meter:
\begin{displaymath}
U\ketwavy\psi_A\ket0_B={\cal V}\ketwavy\psi_A,\quad\forall\;\psi\in H_A.
\end{displaymath}

\noindent Therefore, the isometric property is the condition for the physical
realizability of the transformation in the form  of dynamically reversible
evolution in the bipartite system object--meter.

Index $\alpha$ in the transformation (\ref{V}) accumulates in the classical form an
information associated with the set of initial quantum states $\ketwavy\alpha_A$ of
the object. Values of the index $\alpha$ are mutually uniquely mapped with the set
of classically distinguishable states $\ket\alpha_{AB}$ of the bipartite system.
This correspondence gives us the ground for the definition of the measurement
transformation (\ref{V}) as a sort of purely coherent measurement, which delivers
the output information about the object in the form of entanglement of the state
$\ket\alpha_{AB}$. Such measurement on account of the dequantization effects, which
are pronounced in the partial loss of coherency of the  measurement results without
loss of the classical information, can be represented with the following
superoperator \cite{LLasPhys2004}
\begin{equation}\label{M}
{\cal M}={\cal D}\bigl({\cal V}{\odot}{\cal V}^+\bigr)=\sum_{\alpha\beta}
R_{\alpha\beta}\sqrt{\nu_\alpha\nu_\beta}\ket\alpha_{AB}\brawavy\alpha_A\odot
\ketwavy\beta_A\bra\beta_{AB},
\end{equation}

\noindent where ${\cal D}{=}
\sum_{\alpha\beta}R_{\alpha\beta}\ket\alpha_{AB}\bra\alpha_{AB}\odot\ket
\beta_{AB}\bra\beta_{AB}$ is the dephasing superoperator, $R_{\alpha\beta}$ is
an arbitrary positive-definite matrix with the only diagonal filled with the
unit, which describes the dephasing of the states, and ``$\odot$'' is the
substitution symbol to be replaced with the transformed superoperator (the
density matrix). At $R_{\alpha\beta}\equiv1$, i.e. without dephasing, this
superoperator simply describes the transformation ${\cal V}$ in terms of
density matrix.

General representation of the measurement in the form (\ref{M}) and its purely
coherent modification (\ref{V}) include:
\begin{itemize}
\item The standard projective and entangling measurements \cite{PRA2003} at the
choice of the mapped information in the form of a the complete set of
classically compatible states, the orthogonal basis $\ket k_A$, and as
$\ket\alpha_{AB}$---the duplicated basis $\ket k_A\ket k_B$.

\item The soft measurement \cite{JETP2004} when additionally replacing the orthogonal
basis of the meter $\ket k_B$ with the nonorthogonal set $\ketwavy k_B$.

\item
Considered in this work generalized measurement with partial destruction of the
initial information at the choice $\ket\alpha_{AB}=\ketwavy{e_\alpha}_A
\ket\alpha_B$, where the set of states $\ketwavy{e_\alpha}_A$ is arbitrary and
$\ket\alpha_B$ is formed of orthogonal states and unambiguously maps the values
of the information index $\alpha$, whereas the set $\brawavy\alpha_A$ can count
in nonorthogonal states, as well.
\end{itemize}

The transformation (\ref{V}) corresponding to the generalized measurement takes
the form:
\begin{equation}\label{V0}
{\cal V}=\sum\sqrt{\nu_\alpha}\ketwavy{e_\alpha}_A\ket\alpha_B\brawavy\alpha_A,
\end{equation}

\noindent where in the case of nonorthogonal set $\brawavy\alpha_A$ the
information index $\alpha$ is not unambiguously linked with the classically
distinguishable states of $A$ and its statistics includes the internal quantum
uncertainty of the mapped states $\ketwavy\alpha_A$. It can be formally
interpreted as the number of elementary coherent sub-channel $\ketwavy\alpha_A
\to\ketwavy{e_\alpha}_A \ket\alpha_B$, which links, in general, classically
non-compatible input states of the object $\ketwavy\alpha_A$ with the states of
the bipartite system object--meter $\ketwavy{e_\alpha}_A\ket\alpha_B$.

In case of the soft nondemolition measurement, the orthogonality of the set
$\ketwavy{e_\alpha}_A =\ket k_A$ leads to the unique correspondence with the
informational index $\alpha=k$ and, respectively, to the nondemolition character of
the measurement along the measurable variables of the form $\hat\lambda=
\sum\lambda_k\ket k_A\bra k_A$ and to the complete vanishing of the coherent
information of the meter in respect to the initial state of the object
\cite{JETP2004}.

Nonorthogonality of the set $\ketwavy{e_\alpha}_A$ leads, in its turn, to
reduction of the information remaining in the object, i.e., to the destructive
measurement. During this measurement some coherent information is transferred
into the states of the meter indicator $\alpha$ of which contains quantum
uncertainty with respect to the object states $\ketwavy\alpha_A$ only if the
latter have internal quantum uncertainty and are uniquely represented with the
input states $\ket\alpha_B$ of the meter. In the limiting case
$\ketwavy{e_\alpha}_A\equiv\ketwavy0_A$, the transformation (\ref{V0})
corresponds to the complete transmission of the initial information from $A$
into $B$.

If one uses the orthogonal bases $\ket k_A$ for the sets
$\ketwavy{e_\alpha}_A$, $\ketwavy\alpha_A$, the transformation (\ref{V0})
corresponds to the entirely coherent entangling measurement \cite{PRA2003},
which leads to the equitable probability distribution of the initial
information between $A$ and $B$ and complete absence of the coherent
information about initial states in the subcomponents of the bipartite system
$A+B$.

In general case, distribution of the initial information about the object among
the object and meter is determined by the metric matrix
$Q_{\alpha\beta}{=}\braketwavy{e_\alpha}{e_\beta}_A$ набора
$\ketwavy{e_\alpha}_A$.

In case of the overfull set $\ketwavy{e_\alpha}_A$ the representation of the
operator (\ref{V0}) as a sum over $\alpha$ can be reduced into the
superposition $D^2$ of the projectors by shifting to the minimal orthogonal
basis $\ket k_A$. The respective representation has the form:
\begin{equation}\label{Vkl}
{\cal V}=\sum_{kl}\ket k_A\ketwavy{kl}_B\bra l_A,
\end{equation}

\noindent where $\ketwavy{kl}_B=\sum_\alpha\sqrt{\nu_\alpha}(\alpha\ket l_A\bra
k e_\alpha) \ket\alpha_B$ with the scalar product
\begin{displaymath}
\braketwavy{k'l'}{kl}= \sum_\alpha \nu_\alpha(\alpha\ket l_A\bra k
e_\alpha)_A(e_\alpha\ket{k'}_A\bra{l'}\alpha)_A,
\end{displaymath}

\noindent which is determined only by the states in the Hilbert space of the
object $H_A$. Therefore, the representation (\ref{V0}) clarifies the
transformation of the form (\ref{Vkl}) as setting the correspondence between
the input and output via the classical information index, which surely contains
the internal quantum uncertainty.

\section{Relationship between transformation of the generalized measurement and
its representation in the form of POVM\label{POVM}}

Let us consider the superoperator (\ref{M}), which maps the generalized
transformation (\ref{V0}) taking into account dephasing:
\begin{equation}\label{MM}
{\cal M}=\sum_{\alpha\beta}
R_{\alpha\beta}\sqrt{\nu_\alpha\nu_\beta}\ketwavy{e_\alpha}_A\ket\alpha_B
\brawavy\alpha_A\odot \ketwavy\beta_A\bra\beta_B\brawavy{e_\beta}_A.
\end{equation}

\noindent It corresponds to the probability distribution $P(\alpha)
{=}\bra\alpha_B \hat\rho_B^{}\ket\alpha_B$, where $\hat\rho_B{=}{\rm Tr}_A{\cal
M}\hat\rho_A$, for the results $\alpha$ of the measurement, which are
physically realized in the form of quantum states of the meter. This
distribution has the form
\begin{equation}\label{Pa}
P(\alpha) ={\rm Tr}_A\hat E_\alpha\hat\rho_A
\end{equation}

\noindent with the positive operator valued measure (POVM) $\hat
E_\alpha=\nu_\alpha\ketwavy\alpha_A \brawavy\alpha_A$.

This expression does not depend either on the coherency of the transformation or on
the form of its representation in the output state of the object and corresponding
entanglement in the bipartite system object--meter after the measurement because it
describes only classically compatible information of the object about its initial
state. However, the above expression does not describe the quantum result of the
measurement, but the resulting nonselected information preserved in the object in
quasiclassical form.

The complete resulting information, though displayed in the classically
distinguishable form, is described by the contracted superoperator for the
bipartite system object--meeter
\begin{equation}\label{AtoB}
{\cal M}_B =\sum_{\alpha\beta}
R_{\alpha\beta}\sqrt{\nu_\alpha\nu_\beta}\braketwavy{e_\beta}{e_\alpha}_A\ket\alpha_B
\brawavy\beta_A\odot\ketwavy\beta_A\bra\beta_B,
\end{equation}

\noindent which takes into account quantum correlations with the initial state.
Even for the completely coherent measurement, it contains the decoherence
factor $R^A_{\alpha\beta}= \braketwavy{e_\beta}{e_\alpha}_A$, which is due to
the ignoring of the coherent information bundled in the form of entanglement in
the bipartite system object--meter. This dequantization is not the complete
one. Yet, considering the complete dequantization $R_{\alpha\beta}=
\delta_{\alpha\beta}$ of the output information leads, as one can easily see,
to the transformation
\begin{displaymath}
{\cal M}_B=\sum_\alpha\nu_\alpha\ket\alpha_B\brawavy\alpha_A\odot
\ketwavy\alpha_A\bra\alpha_B.
\end{displaymath}

\noindent The respected probability distribution for this transformation is
given by Eq.\ (\ref{Pa}) on the algebra of classical events described with the
set of compatible states $\ket\alpha_B$ and its subsets.

It is worth to note here that the generalized measurements in terms of POVM
have been widely discussed in the literature, particularly, in connection with
the problem of optimal measurement of continual quantum variables, i.e.,
coordinates and momenta \cite{helstrom,grishanin,kholevo}. However, our
consideration is qualitatively different because we consider the
finite-dimensional Hilbert space $H_A$ and, respectively, discuss potentially
information about all quantum states in this space, whereas in the previous
works the analysis has been done for the infinite-dimensional space, which
cannot be applied to our case.

\section{Selected generalized measurement\label{selective}}

A special case of the generalized measurement is the selected measurement,
which has different from the entangling measurement generalized set of the
output states of the corrected object: $\ket k_A\to
\ketwavy{e_\alpha}_A=\ketwavy{k}_A$, $k=1,\dots,D$. These states are different
from the basis states $\ket k_A$ of the measurable variable and, being in
general case represented by the nonorthogonal set of states, prevent the object
preserving the initial state of the measurable variable.

In case of purely coherent measurement, the meter attains a nonzero coherent
information about the initial state of the object, which in the limiting case of
$\ketwavy{k}_A\equiv\ket0_A$ is the complete information, i.e., the information
equal numerically to the initial entropy of the object. In this case, the
information relationships for the mapping object--meter reproduce obviously the same
relationships for the mapping object--object for the case of the soft measurement,
transformation for which is described with the transformation $H_A\leftrightarrows
H_B$ of the resulting states of the object and meter. Therefore, the respective
dependencies given in \cite{JETP2004} for the coherent information object--object
for the two-level system, retain their validity in the present case, as well.
Quasiclassical information attained by the meter is due to the absolute accuracy of
the measurement always complete, i.e., the amount of information coincides with the
entropy of the measurable variable.

\section{Nonselected generalized measurement\label{nonselective}}

The nonselected measurement gives us another special case of the measurements
for which the set of mapped states $\ketwavy\alpha_A$ includes {\em all}
quantum states of the object. In this case, the information index $\alpha$
unambiguously maps all physically distinguishable elements of the Hilbert space
$H_A$ and the appropriate representation of the set of its states is the unit
$(2D-2)$-mensional sphere of the real Euclidean space.

\begin{figure}
\begin{center}
\epsfysize=0.3\textwidth\epsfclipon\leavevmode\epsffile{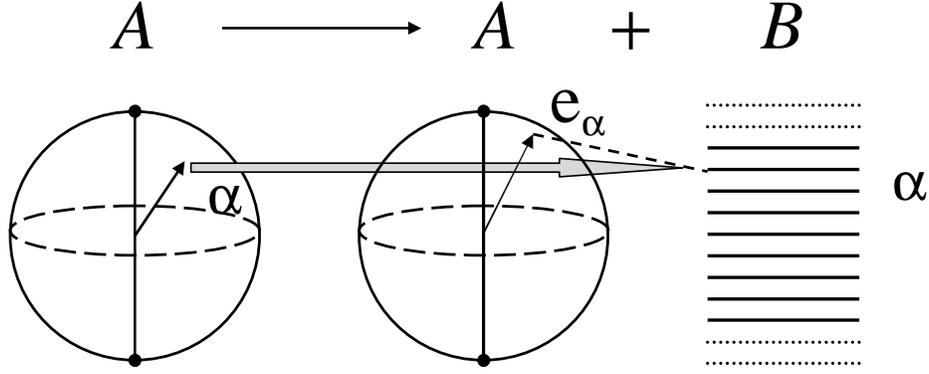}
\end{center}
\caption{Mapping of the elementary states in the process of nonselected generalized
measurement. New states $\ketwavy{e_\alpha}_A$ of the object are, generally,
different from $\ketwavy\alpha_A$.\label{fig1}}
\end{figure}

Then, the respected generalized measurement is the map $H_A\to H_A\otimes H_B$
with the states of the meter $H_B=L_2(H_A)$ with the wave functions of the
continual argument $\psi_B^{}(\alpha)$. Multiplicity of the states
$d\nu=\sum_{dV}\nu_\alpha$, which corresponds to the element in the set
$\alpha\in dV$, has in this case the form $d\nu=DdV/V$, where $V$ is the entire
volume of the hypersphere of the physical states. Fig.~\ref{fig1} illustrates
an example of the two-level system.

\subsection{Distribution of information between object and meter\label{distribution}}

The amount of information preserved in the object is determined by the
information capacity of the overfull basis $\ketwavy{e_\alpha}_A$, which
duplicate information that is represented by the states of the meter
$\ket\alpha_B$, but, in general case, corresponds to a partial or complete loss
of the initial information $\ketwavy\alpha_A$ of the object. In case of
completely coherent measurement, i.e. at $R_{\alpha\beta}\equiv1$, the
information capacity of the basis $\ketwavy{e_\alpha}_A$ for the pure input
state $\hat\rho_A=\ket\psi_A\bra\psi_A$ is determined by the entanglement
$E[\ket{\psi}_{AB}]$ of the resulted state $\ket{\psi}_{AB}={\cal V}\ket\psi_A$
of the bipartite object--meter system. The meter in this case contains all
\emph{accessible} information \cite{fuchs} about all Hilbert space of object
states, which is represented, however, in quantum form including the
entanglement with the object. This information is reduced into classical form
either after additional projective measurement or after entirely dephasing
transformation $\cal D$ в (\ref{M}) at $R_{\alpha\beta}=\delta_{\alpha\beta}$,
which are equivalent from information point of view.

Let us illustrate the distribution of information among the object and meter on
an example of a two-level system with $D=2$ using as $\ketwavy{e_\alpha}_A$ all
states of a part of the Bloch sphere, which is formed by the mapping
$\vartheta\to q\vartheta$, where $0\leqq q\leqq1$ is the compression
coefficient of the initial Bloch sphere that is mapped into its part
corresponding to $0\leqq\vartheta\leqq\pi q$. With this choice of the mapping,
at $q<1$ there is some asymmetry with respect to the value of the polar angle
$s$ of the initial state $\alpha_0=(s,\varphi_0)$. This asymmetry reaches its
maximum at $q=0$ and vanishes at $q=1$.

\begin{figure}
\begin{center}
\epsfysize=0.4\textwidth\epsfclipon\leavevmode\epsffile{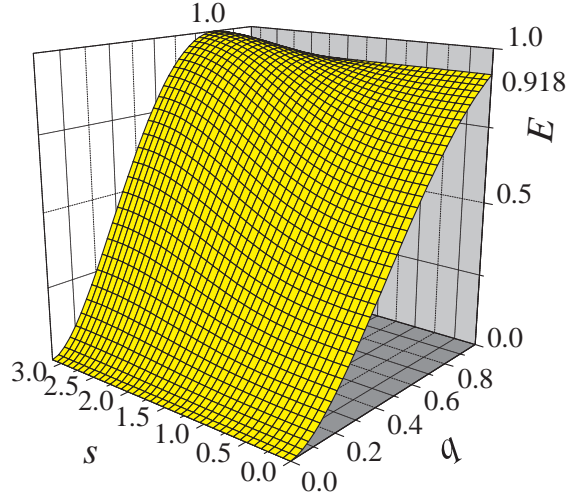}
\end{center}\caption{The degree of entanglement $E$ (in bits) versus the compression
coefficient $q$ of the Bloch sphere and the angle $\vartheta=s$ of the initial
state. The maximum value $E=1$~bit is achieved at $s=\pi$ and $q=0.7978$. At
$q=1$ the degree of entanglement does not depend on $s$ and is equal to the
entropy $E_0=0.918$~bit for the pure state of the qubit after its complete
depolarization.\label{fig2}}
\end{figure}

The entanglement in the object--meter system that arises after the measurement
can be written as the entropy $S[\hat\rho'_A]=-{\rm Tr}\,\hat\rho'_A\log_2 \hat
\rho'_A$ of the partial density matrix $\hat\rho'_A={\rm Tr}_B\ket\psi_{AB}\bra
\psi_{AB}$ of the transformed object state \footnote{While calculating
$\hat\rho'_A$, the calculation of the trace over $B$ in the considered
continual limit is reduced to the corresponding integral on the Bloch sphere
with the differential $d\nu=\sin\vartheta d\vartheta d\varphi/2\pi$.}. The
corresponding dependence $E(s,q)$ is shown in Fig.~\ref{fig2}.

The results of the analysis of the information distribution in the two-level system
for the completely nonselected representation of the final state of the object at
$q=1$ and, respectively, $\ketwavy{e_\alpha}_A=\ketwavy\alpha_A$ is an obvious one,
even without any calculations, because in this case $\hat\rho'_A$ corresponds to the
entirely depolarized initial state (see, for example, Eq.~(3.115) at $p=1$ in
Ref.~\onlinecite{preskill})
\begin{displaymath}
\hat\rho'_A=(2/3)\ket{\alpha_0}
\bra{\alpha_0}+(1/3)\ket{\overline\alpha_0}\bra{\overline\alpha_0}
\end{displaymath}

\noindent ($\ket{\overline\alpha_0}$ is orthogonal to $\ket{\alpha_0}$) and,
independently from $\alpha_0$, $E=E_0=(2/3)\log_2(3/2)+(1/3)\log_2(3/1)$.

However, the result $E=1$~bit, i.e., the complete entanglement between the object
and the meter, which is achieved at the orientation of the initial state $s=\pi$,
opposite to the Bloch sphere compression point $\vartheta=0$, and at the
intermediate value of the compression coefficient, is not trivial and requires a
qualitative elucidation.

We can do that easily because at the chosen orientation the problem is
symmetrical in respect to the axis of the Bloch sphere, thus the density matrix
in the respective basis is a diagonal one and has the form:
$\hat\rho'_A=p_1\ket1_A\bra1_A+p_2\ket2_A\bra2_A$. Also, the direction
$\vartheta=0$ is opposite to the direction of the initial state $s=\pi$ and,
therefore, the probability $p_1$ in accordance with the given above equation
for $q=1$ is simply $p_1=1/3$. If one changes the compression coefficient up to
the value of $q=0$, which corresponds to the collapse of the Bloch sphere into
the point $\vartheta=0$, the probability of the opposite state $p_2$ reduces up
to the zero and, respectively, the probability $p_1$ grows up to the unit in
accordance with the following analytical formula:
\begin{displaymath}
p_1=1-p_2=\frac{3- 2q^2+\cos\pi q}{4(1-q^2)}+\frac{1-\cos\pi q}{4(4-q^2)}.
\end{displaymath}

\noindent This probability due to it continuity passes the value of $p_1=1/2$,
which corresponds to the maximum possible entanglement between two systems, one
of which if the two-level system (qubit).

Note also that the maximal degree of entanglement $E_0=0.918$~bit, achieved at
the exact reproduction by the object after the measurement of all states of the
Hilbert space, is very close to the maximal entanglement, which is achieved at
the totally coherent nondemolition entangling measurement. This, however, can
be achieved only with the optimal choice of the initial wave function of the
object. In case of the completely nonselected measurement, the degree of
entanglement is invariant with regard to the initial state
$\ketwavy{\alpha_0}_A$ because all the states are due entirely equal.

\section{Competition between object and meter in selection of the nonselected quantum
information \label{concurr}}

In case of nondemolition quantum measurement, there is no competition between
the object and the meter because classically compatible information retrieved
at such measurement can be duplicated without bound. However, with the choice
of nonselected information, which is connected with the nonorthogonal overfull
set $\ketwavy\alpha_A$, typically, for instance for the quantum key
distribution protocols \cite{BB84}, the competition arises. It is due to the
impossibility of nondemolition duplication of the information about the
nonorthogonal quantum states. Mathematically, such competition can be
sufficiently treated with the Holevo information \cite{kholevo}, which
implicitly takes into account quantum nature of the information.

The Holevo information is defined for the semiclassical channel, which is
characterized by the density matrix $\hat\rho(\alpha)$ depending on the
classical messages $\alpha$ at the input of the channel, as
\begin{equation}\label{Ih}
I_h=S[\hat\rho]-\int P(d\alpha)S[\hat\rho(\alpha)],\quad
\hat\rho=\int\hat\rho(\alpha) P(d\alpha),
\end{equation}

\noindent where $P(d\alpha)$ defines the probability distribution or the frequencies
of the classical messages $\alpha$. On can easily see that in the considered
transformation of the quantum measurement the classical parameter $\alpha$
corresponds to the informational index of the initial states of the object
$\ketwavy\alpha_A$ and two considered channels, object--object and object--meter,
are described with the averaging of the wave function ${\cal V}\ketwavy\alpha_A$ of
the combined object+meter system (or, in general, of the density matrix, which
results after the incoherent transformation (\ref{M})) over the competing system.
For the uniform distribution $P(d\alpha)$, the density matrices for the
corresponding channels have the following form:
\begin{eqnarray}
\hat\rho_A(\alpha)&=&\frac{D}{V}\int dV_\beta\,|\braketwavy\beta\alpha_A|^2
\ketwavy{e_\beta}_A\brawavy{e_\beta}_A\,,\quad\hat\rho_A=\frac{1}{V}\int dV_\beta
\ketwavy{e_\beta}\brawavy{e_\beta}\,;  \label{rhoAa} \\
\hat\rho_B(\alpha)&=&\sum_{\beta\beta'}\sqrt{\nu_\beta^{}\nu_{\beta'}}
\braketwavy{e_\beta}{e_{\beta'}}_A\braketwavy{\beta'}\alpha_A\braketwavy
{\alpha}{\beta}_A\ket{\beta'}_B\bra{\beta}_B\,; \\
\hat\rho_B&=&\frac{1}{D}\sum_{\beta\beta'}\sqrt{\nu_\beta^{}\nu_{\beta'}}
\braketwavy{e_\beta}{e_{\beta'}}_A\braketwavy{\beta'}{\beta}_A\ket{\beta'}_B
\bra{\beta}_B\quad\to\\ \hat{\tilde\rho}_B&=&\frac{1}{V}\int dV_\beta\ketwavy
\beta_A\ketwavy{e_\beta^*} \brawavy{e_\beta^*}\brawavy\beta_A\,. \label{rhoB}
\end{eqnarray}

\begin{figure}
\begin{center}
\epsfysize=0.4\textwidth\epsfclipon\leavevmode\epsffile{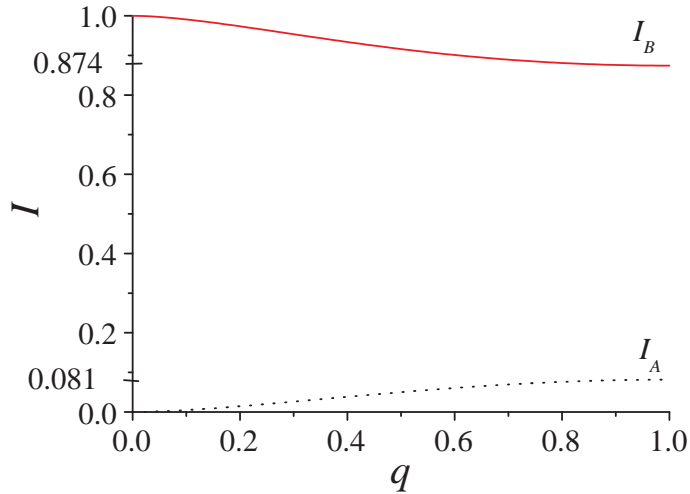}
\end{center}\caption{The measured amount of Holevo information ($I_B$) and
information preserved in the object ($I_A$) about the equally distributed
ensemble of initial states $\ketwavy\alpha_A$ of the object (qubit) versus the
degree $q$ of the preserved information in the object. The maximum amount of
information about the object $I_A=0.081$~bit corresponds to the minimum of
measured information $I_B=0.874$~bit.\label{fig3}}
\end{figure}

\noindent The latter equation is the isometric display of the continual density
matrix of the meter into the discrete space $H_A\otimes H_A$, which realizes
the active subspace of the states and that is used for the numerical
calculations. Entropies of the density matrices $\hat\rho_A(\alpha)$,
$\hat\rho_B(\alpha)$ coincide with each other, so that there is no need in
using the continual representation. Respective dependencies for the information
(\ref{Ih}) of the meter and the object that are calculated with the help of
Eqs.\ (\ref{rhoAa}) and (\ref{rhoB}) are shown in Fig.\ \ref{fig3}. They
demonstrate the relatively weak competition character, for example, by contrast
with the competition of the coherent information during the selective
measurement, when preserving the entire information in the object corresponds
to its complete absence in the meter \cite{PRA2003}.

\section{Conclusions}

In conclusion, the class of the partially-destructive quantum measurements
discussed in this paper generalizes using the common mathematical formalism all
the key classes of the quantum measurements discussed in the literature:
standard projective, entangling, soft, destructive, coherent, and dequantized
measurements.

The nonselective subclass of the partially-destructive quantum measurements
has, as it was demonstrated on example of the two-level system, an interesting
feature: the degree of entanglement in the bipartite object--meter system can
reach its maximum value at the intermediate degree of preserving in the object
its initial information. When the object preserves the maximum possible
information about its initial state, the degree of entanglement nearly achieves
its maximum for all possible pure initial states. It has also been shown that
the nonselective quantum measurements realize the equitable measurement of all
the dynamical variable of the measurable quantum system. Therefore, they are
characterized with the essentially low level of competition of quantum
information in between the object and the meter, by contrast with the totally
selective measurements.

\acknowledgements

This work was partially supported by RFBR grant No.04--02--17554 and INTAS
grant INFO 00--479.


\begin{thebibliography}{99}

\bibitem{bowmeester}
{\em The Physics of Quantum Information: Quantum Cryptography, Quantum
Teleportation, Quantum Computation}, edited by D. Bouwmeester, A. Ekert, and A.
Zeilinger, (Springer-Verlag, New York, 2000).

\bibitem{Neumann}
J. von Neumann, \emph{Mathematical Foundation of Quantum Mechanics} (Princeton
University Press, Princeton, 1955).

\bibitem{sudbery}
A. Sudbery, \emph{Quantum Mechanics and the Particles of Nature} (Cambridge
Univ. Press, New York, 1986).

\bibitem{PRA2003}
B. A. Grishanin and V. N. Zadkov, {\em Phys. Rev.} {\bf A 68},
022309 (2003).

\bibitem{JETP2004}
B. A. Grishanin and V. N. Zadkov, submitted to Sov. JETPH.

\bibitem{mensky}
M. B. Mensky, \emph{Quantum measurements and decoherence: models and phenomenology}
(Kluwer Academic: Dordrecht, 2000).

\bibitem{RTE02}
B. A. Grishanin and V. N. Zadkov, J. Commun. Technology and Electronics
\textbf{47}, 933 (2002).

\bibitem{LLasPhys2004}
B. A. Grishanin and V. N. Zadkov, \emph{Laser Phys. Lett.} {\bf
2}, No. 2 , 106 (2005) / DOI 10.1002/lapl.200410136.

\bibitem{helstrom}
C. W. Helstrom, \emph{Quantum detection and estimation theory} (Academic Press,
New York, 1976).

\bibitem{grishanin}
B. A. Grishanin, Izv. Akad. Nauk SSSR, Ser. Tekh. Kiber. \textbf{11}, 127
(1973); LANL e-print quant-ph/0301159 (2003).

\bibitem{kholevo} A. S. Holevo, \emph{Probabilistic and Statistical Aspects of Quantum Theory}
(North Holland, 1982).

\bibitem{fuchs}
C. M. Caves and C. A. Fuchs, LANL e-print quant-ph/9601025 (1996).

\bibitem{preskill}
J.~Preskill, http://www.theory.caltech.edu/people/preskill/ph229/, p. 106.

\bibitem{BB84}
Ch. H. Bennett and G. Brassard, in {\em Proceedings of the IEEE International
Conference on Computer, System and Signal Processing, Bangalore, India} (IEEE,
New York, 1984), p. 175.

\end{thebibliography}
\end{document}